# In situ resonant photoemission and X-ray absorption study of the BiFeO$_3$ thin film

Abduleziz Ablat[a], Mamatrishat Mamat [a*], Yasin Ghupur[b], Rong Wu[b], Emin Muhemmed[b], Jiaou Wang[b], Haijie Qian[b], Rui Wu[b], Kurash Ibrahim[b*]

[a]*School of physical Science and Technology, Xinjiang University, Urumqi 830046 China*
[b]*Beijing Synchrotron Radiation Facility, Institute of High Energy Physics, CAS, Beijing 100049, China*

## Abstract

Multiferroic bismuth ferrite (BiFeO$_3$) thin films were prepared by pulsed laser deposition (PLD) technique. Electronic structures of the film have been studied by *in situ* photoemission spectroscopy (PES) and x-ray absorption spectroscopy (XAS). Both the Fe 2*p* PES and XAS spectra show that Fe ion is formally in +3 valence state. The Fe 2*p* and O *K* edge XAS spectra indicate that the oxygen octahedral crystal ligand field splits the unoccupied Fe 3*d* state to t$_{2g}$↓ and e$_g$↓ states. Valence band Fe 2*p*-3*d* resonant photoemission results indicate that hybridization between Fe 3*d* and O 2*p* plays important role in the multiferroic BiFeO$_3$ thin films.

## 1. Introduction

Multiferroics are a group of materials that a same system simultaneously having different forms of ferroic properties such as ferroelectricity, ferromagnetism, and ferroelasticity [1]. In multiferroics, magnetization can be tuned by applied electric field, and electron polarization by applied magnetic field. These effects called magnetoelectric effect[2]. Few multiferroic materials, YMnO$_3$, PbVO$_3$, BiCrO$_3$, BiMnO$_3$ and BiFeO$_3$ (BFO), exhibit as of natural occurring phases [3-5]. Among them, the BFO is the only material that shows magnetoelectric effects at room temperature[6]. The BFO is considered to be a promising candidate for electronic device application benefitting from its high Curie ($T_C$~825 °C) and Néel ($T_N$~370 °C) phase transition temperatures[4]. Compared to its bulk form，BFO thin film shows apparently improved ferroelectric and magnetization characters[7].

[*] corresponding author
E-mail Address: kurash@ihep.ac.cn (K. Ibrahim)
mmtrxt@xju.edu.cn (M.Mamat)

Understanding the multiferroic properties of BFO in terms of its electronic structure is important, because both the electric and magnetic properties closely relate to it. An open problem with BFO is that it loses easily the ferroic characters with high current leakage. The current leakage was attributed to small amount of Fe$^{2+}$ ion or oxygen vacancies. To solve this problem, there has been investigation with different growth techniques and various doping methods [8-10]. Decrease or elimination of the Fe$^{2+}$ ion impurity is usually judged through the trends of decreased current leakage and enhanced ferroelectric hysteresis loops. Experimental investigations in terms of valence band electronic structure are rare [11,12]. On the other hand, results in the electronic band structure of BFO by theoretical calculation are controversial [13,14]. Density of state analysis [13] shows that the Bi 6*p* state locates above 4eV of the gap, and the origin of the ferroelectricity is ascribed as O 2*p*–Bi 6*p* dynamic hybridization. Other studies [14] show the Bi 6*p* state has a contribution on both valence and conduction bands through hybridization with both Fe 3*d* and O 2*p* states. In the current work we aim at the investigation of the valence band states of the BFO by means of photoemission method to uncover possible contributing routes in regard to the controversial conclusions.

We first prepared BFO thin films by PLD method for *in situ* PES and XAS measurement. The advantage of thin film over bulk state is at least threefold. Firstly, thin film can free from charging effect by controlling film thickness that is a main unwanted factor in the PES measurements. Secondly, thin films can be grown by thickness controlling, as well as with slightly varied stoichiometry by regulating ambient oxygen partial pressure. In the last, the *in situ* prepared thin films can free from surface contamination that is crucial for PES measurements, due to the measurement itself is highly surface sensitive. The results show the films prepared under higher oxygen partial pressure is near to stoichiometry, and the Fe ion is formally in valence-three state. Oxygen *K* edge XAS and valence band resonant photoemission spectroscopy (RPES) through Fe 2*p* core level excitation indicate the conduction and valence bands are mainly of Fe 3*d* and O 2*p* hybridization state.

## 2. Experiment

Ceramic BFO target of 20 mm diameter with 2 mm thickness has been prepared by sintering 1.01:1 mixtures of $Bi_2O_3$ (99.99 at.%) and $Fe_2O_3$ (99.99 at.%) at 820 °C, following the routine way of solid state reactions. Slight excess of $Bi_2O_3$ has the role to compensate the preferential loss of bismuth during sintering. X-ray diffraction (XRD) result of the target material indicates that the sample is in single phase of BFO. The BFO films are prepared by PLD method on the Pt (111)/$TiO_2$/$SiO_2$/Si (100) and quartz substrates in a chamber connected to the photoemission system at 4B9B beam line of Beijing Synchrotron Radiation Facility [15].

Prior to deposition, the Pt (111)/$TiO_2$/$SiO_2$/Si (100) substrates were preheated at 700°C to eliminate surface contaminations. To grow near stoichiometric BFO films, the substrates were kept at 550°C and the films were grown at different oxygen partial ambient pressures. The laser fluence was 2.3J/cm$^2$ with pulse repetition rate 1.5Hz. A stoichiometric film was obtained at oxygen partial pressure $P_{O_2}$=5.6 Pa, consistent with reported results [16,17]. After deposition, the sample was transferred to the PES chamber under a background pressure of ~$10^{-8}$ Pa. The overall energy resolution was 0.2–0.7 eV, depending on the selected monochromatic photon energies. Photoemission spectra were calibrated to the in-substrate Pt $4f_{7/2}$ peak at 71.2 eV. The XAS measurement carried out with total electron yield detection mode.

## 3. Results and Discussion

*3.1 PES characterization of the films*

Usually, PES measurement can be used to check whether a sample is stoichiometric or not and to estimate relative atomic ratios in the system. Determination of ionic valence state is through indirect way by measuring energy positions of concerned elements' main peaks and related satellite peaks, by inspecting the ratios of peaks intensity of constitute atoms. Fig.1 illustrates a wide range scan PES spectrum of the BFO film at 900 eV that gives overall status information about the constituent elements of the system. The dominating signals are of core-level photoemission peaks and Auger lines of Bi, Fe, and O atoms. Absence of the C 1$s$ signal at about 285 eV binding energy indicates that the surface is free from contamination, thanks to the *in situ* operations. Characteristic core-level photoemission peaks of Fe 2$p_{3/2}$ 711.3 eV, Fe 2$p_{1/2}$ 725.4 eV, Bi 4$p_{3/2}$ 680.7 eV, Bi 4$d_{5/2}$ 441.4 eV, Bi 4$d_{3/2}$ 465.7 eV, Bi 4$f_{7/2}$ 159.3 eV, Bi 4$f_{5/2}$ 164.6 eV, O 1$s$ 530.3 eV and Pt 4$f$ at 71.2 eV from substrate, and Auger decay channels of Fe $L_3M_{23}M_{23}$, Fe $L_3M_{23}M_{45}$, Fe $L_3M_{45}M_{45}$, O KVV are labeled on the spectrum.

The O 1$s$ narrow range scan spectrum (Fig.2 a) with a higher energy resolution and better statistics show asymmetric peaks near 530 eV. The spectrum can be fitted by two peaks locating at lower and higher binding energies (LBE and HBE), respectively. The LBE peak is attributed to the O 1s of perfect BFO phase, while the HBE one relates with oxygen defects in samples [18]. The peak area of the LBE is larger than that of HBE, indicating the sample was in low oxygen vacancy, the BFO is mainly with perfect stoichiometric constituent.

The Fe 2p core level photoemission spectrum (Fig.2 b) subjects to be applied to determine the valence state of Fe ions, where it splits into Fe $2p_{3/2}$ at 711.3 eV and Fe $2p_{1/2}$ at 725.4 eV as result of spin–orbit coupling effects. The Fe $2p_{3/2}$ peak shape resulted in fitting is normally inspected to determine if the cation is in uni-valence or in multi-valence state, where the Fe $2p_{3/2}$ is decomposed into symmetric superposition components of $Fe^{2+}$ and $Fe^{3+}$[19]. But recent literature report suggests that decomposing the Fe 2p into symmetric components is not feasible [20]. The way to fix the formal valence state is by satellite peak of $Fe^{3+}$ and $Fe^{2+}$ in PES, because of different d orbital electron configurations, relaxation of the $Fe^{2+}$ and $Fe^{3+}$ show satellite peaks at 6 eV and 8 eV, respectively, above their $2p_{3/2}$ main peaks [16]. The measured spectrum in Fig.2(b), a satellite peak 8eV above the Fe $2p_{3/2}$ 711.3 eV, confirms that the iron atom is formally in $Fe^{3+}$ state. In the limits of the PES measurement sensitivity to molar portion in system, there is no apparent evidence showing up for the existence of $Fe^{2+}$ ion in the underlying BFO thin film.

The Bi $4f_{7/2}$ and $4f_{5/2}$ spin-orbit coupling components locating at 159.3 eV and 164.6 eV respectively, in Fig.2 (c), are consistent with literature values of $Bi^{3+}$[21]. There are no shoulders appearing either at higher or lower binding energy sides as being clues for the existence of higher than $Bi^{3+}$ oxidized or lower than $Bi^{3+}$ reduced metallic bismuth states[22]. The estimated molar ratio of Bi to Fe from the peak area of the Bi 4f and Fe 2p, which are normalized to respective atomic orbital cross sections at the specific excitation energy[23], is near to 1:1 indicates that composition of the thin film under proper $P_{O_2}$ pressure maintained almost the same with its target precursor.

*3.2. O K edge XAS and Valance band photoemission*

The XAS provides information on the excitation of a core electron into unoccupied states as function of photon energy, that is about a cross sectional variation of measured photoelectron DOS versus photon energy. It implies unoccupied density of state and crystal field splitting effect, and most importantly provide signature on hybridization between atoms.

The left and right panels of Fig.3 show the valence band PES and O *K* XAS spectra of the BFO thin film. Valence band PES represents the occupied DOS distribution below Fermi level. In Fig 3(a) the VB spectrum can be divided into two main block structures labeled as α, β, γ, δ from Fermi edge to below ~9 eV and a broad peak at ~12 eV. The feature around 12 eV originates from Bi 6s states according to the DOS result by first principle calculation [24]. In the coming section, we will discuss an enhanced feature of the four α, β, γ, δ structures by resonant photoemission at Fe $L$ edge.

The O $K$-edge XAS spectra in Fig 3(b) provides information on the hybridization of O $2p$ with Fe $3d$ states. Under ionic picture, the O $2p$ is considered fulfilled with six electrons. Then following the definition of XAS, we should observe any absorption effect at the O 1s region, but actually we do observe a copious structure in this region. The observed results imply a certain portion of empty O $2p$ orbitals with nontrivial probability due to hybridization between the Fe $3d$ [25], the hybridization results in dynamic charge transfer from O $2p$ to Fe $3d$. The oxygen $K$ edge further manifests an extended structure up to at least 17 eV above the threshold, implying an electronic structure resulted in various effects such as ligand field splitting, hybridization of the oxygen $2p$ and transition metal $3d$ states. The region near to the absorption edge, labeled as $a'$, $a$ and $b$, the most sensitive region to those effects, is expanded and shown as inset in Fig.3. Among them the features $a'$ and $a$ have the key importance for understanding inter-atomic interactions. The $a'$ at ~530.8 eV is attributed to the O $2p$+Fe $3d$ $t_{2g}\downarrow$ characteristic states, the peak $a$ to the O $2p$ + Fe $3d$ $e_g\downarrow$ states, where $\downarrow$ denotes the minority spin states.

Reason for how that O $2p$ + Fe $3d$ hybridization leads to the status quo $a'$, $a$ structure is twofold. Firstly, the XAS is a local process in which an electron is promoted to an unoccupied electronic state, which couples to the original core level restricted by the electric dipole selection rule by parity consideration, which states that the change in the angular momentum quantum number can only take $\Delta L=\pm 1$ values, while the spin keeps unchanged [26]. For the excitation of an O 1s electron of $l=0$, the $\Delta L=\pm 1$ requirement means that only O 2p character $l=1$ can be reached, and an inter-atomic O 1s—Fe 3d transition is not allowed. In terms of ionic picture for oxygen in oxides, this prohibits, theoretically, occurrence of O 1s—O 2p absorption. These apparent idealized situations do not hold regarding the experimental reality, by showing up a wide peak envelope in 529~535 eV region with fine internal structures as of complicated interaction effects. These experimental facts imply that the rigid ionic picture is not appropriate for understanding the observed results, and the results reveal that there exist strong interactions of O ions with the central transition metal Fe ions. The interactions open up channels available for the transition between O 1s and Fe 3d-like orbitals. Such interaction (hybridization) brings about transfer of certain amount of electron density, i.e., dynamic charge transfer from O 2p to Fe 3d orbital that creates hole-state in the O 2p. The hole-state creation on the ligand side is instrumental for understanding the experimental features observed at the O K edge. Secondly, the observed ~1.4 eV distance between the $a′$ and $a$ peaks in the O 1s XAS, a similar value to the crystal-field splitting 10Dq induced between the two empty states $t_{2g}\downarrow$ and $e_g\downarrow$, is almost the same as observed in the Fe 2p XAS spectra in Fig.4. These imply that they are of Fe 3d character and provide a solid base for the concept of dynamic charge transfer through hybridization.

A common point for the peak *b* is to ascribe it as of Bi 6*sp* character[14], that is due to interaction between the full filled O 2*p* and partial occupied Bi 6*sp* states allow to form a certain degree of covalent bonding between the O 2*p* and Bi 6*sp*. There exists also controversial points of view regarding the *b*'s origin [27], where refers the peak *b* to as result of the interactions between the unoccupied O 2*p* and Bi 5*d* states. The peak *c* at ~542 eV derives from hybridization of O 2*p* and Fe 4*sp*[28].

*3.3. Fe 2p XAS and Fe 2p-3d resonant photoemission*

Fig.4 shows the Fe 2*p* XAS spectrum of BFO thin film together with those of reference oxides of formal trivalent $Fe_2O_3$($Fe^{3+}$), divalent FeO ($Fe^{2+}$) [29], and plus the atomic multiplet calculation result. The XAS data of $Fe_2O_3$ and FeO were shifted by ~3 eV to align with the measured one. Calculation performed with atomic multiplet model in octahedral symmetry for different 10Dq values varying from 1.0 to 1.8eV, a multiplet splitting energy range 10Dq suited for the 3d transition metal 2*p* edges [30]. The Slater integrals were scaled to 80% of the associated atomic values. For simulate lifetime effects and instrument resolution, discrete multiplets were broadened with a Lorentzian of 0.3 eV and a Gaussian of 0.3 eV at both $L_3$ and $L_2$ edges. A good agreement achieved between the calculation and experiment for a value at 10Dq~1.6 eV.

The measured energy separation between $t_{2g}\downarrow$ and $e_g\downarrow$ states, both for $L_3$ and $L_2$ levels, about 1.4 eV corresponds to the crystal field splitting energy 10Dq. It leads the Fe $3d^5$ electrons to high-spin configuration $t_{2g}^3 e_g^2$ ($^6A_{1g}$) [31]. Resemblance of the BFO line shapes to that of $Fe_2O_3$, and a significant deviation from FeO indicate the Fe ions are formally in trivalent ($Fe^{3+}$:$3d^5$) state, confirming the evidence drawn from the PES.

Resonant photoemission (RPES) spectra shown in Fig.5(a) recorded between photon energy 702 eV and 718 eV spanning the Fe $2p_{3/2}$ threshold region of XAS in Fig.4 to look at the valence electronic structure relation of Fe with its surrounding ligand O atoms. The most resonantly enhanced spectrum at 710.5 eV and two off-resonance spectra below and above the threshold are shown in Fig.5(b) to contrast resonant and off-resonant effects. The spectra are normalized to the storage ring current and the spectrum sweeping numbers. The valence band spectra consist of four main structure labeled as α, β, γ and δ, the same as those observed in Fig.3(a).

An overall fuzzy enhancement in spectral intensity of the peaks and then decreasing when the excitation energy pass through the Fe $2p_{3/2}$ threshold region is clear in the spectra shown in Fig.5(b), that reflect the sensitiveness of the Fe 3*d* characteristic partial DOS to the threshold excitation. A fuzzy broad block of peaks, which simultaneously varies all together in lack of independently enhanced peak shapes below 0-14 eV Fermi level, reaches maximum at least with a factor of ten at excitation energy 710.5 eV relative to those excited below and above at this energy. The valence band peaks photoemission intensity increase with such fuzziness as whole, instead of observing a specific peak enhancement, reveals at least two kinds of information. The first is that the photoemission in the on-resonance region has apparently additional contribution source from that of photoemission in the off-resonance region. The second remarkable point reveals that the resonantly enhanced intensity shape of valence bands strongly relates to surviving environment of Fe 3*d* electrons, instead of the Fe 3*d* partial density of state *per se*.

The resonant photoemission enhancement in an order of magnitude at threshold excitation is a generic situation for most elements in condensed matter systems, and its origin can differ for different systems. One can say at least one common cause about the significant difference between the off-resonance and on-resonance is that the former mainly results in the single-photon single-channel process, and the latter through single-photon multiple-channel processes. Here the single-photon is referred to as monochromatized single energetic photons, single-channel process means the measured DOS corresponds to single-photon single-channel (or direct) photoemission process. While the single-photon multiple-channel (or indirect) processes make an additional aggregative contribution from multiple-channel de-excitations against single photon excitation, which are in addition to the DOS results in the single-photon single-channel process.

The above discussed processes are schematically shown in Fig.6. Among them, the single-photon single-channel process refers to as the direct photoemission ①. In this process, whenever the photon energy is below or above the threshold region, the measured intensity of the valence band comes only contribution from direct photoionization of the valence state electrons, no other processes are countable for it. When the excitation photon energy match with the potential energy of certain inner shell, the resonantly excited electron from that shell leads to the processes ② or ③. These last two processes are indirect photoemissions, their contributions add up to that of the normal process ①, at the end they give the measured results where the intensity has an enormous increase. The single-photon multiple-channel processes at the threshold region are actually the summation of ① + ② + ③ + ⋯. Here in the both processes ② and ③, the neutral initial state $[2p^6 vb 3d^n]$, where the $vb$ refers to all the rest valence band electrons except that of Fe $3d^n$, the Fe 2p state electron absorbs a photon and excites into intermediate

$[2p^5vb3d^{n+1}]^*$ or $[2p^53s^2vb3d^{n+1}]^*$ excited neutral states by creating a primary vacancy in the Fe $2p^5$ core shell and promoting one electron to $vb3d^{n+1}$ state. These states then de-excite through corresponding channels into respective final states. The $[2p^5vb3d^{n+1}]^*$ state de-excites into $[2p^6vb3d^{n-1}]^-$ final state through super-Coster-Kronig channel [32], and the $[2p^53s^2vb3d^{n+1}]^*$ state into $[2p^63s^1vb3d^n]^-$ through Coster-Kronig process [33]. These are the main origins for the observed intensity increase at threshold excitations in photoemission measurement.

Unravel electronic structural origin of the density of state enhancement as a whole with the threshold excitation in the valence band region, where the valence band states marked as α, β, γ and δ, requires to understand the relationship between Fe $3d^n$ and the rest $vb$ electrons in the region. In other words, need to have an appreciation of Fe $3d^n$ surviving environment. The resonant valence band photoemission measured by excitation of inner-shell electron which leaves a core-hole behind play the role as of pump-probe, by which one is able to inspect the electronic structural interaction of constituent elements in the valence region. As shown in Fig.5(a), the four feature through enhanced in order scale, they are with as whole but without a single peak with apparent sharpening, seemingly if they were a single body. It implies there have exist a strong hybridization between Fe 3d with its surrounding environment, such as O 2p and Bi 6p states. One can approximately assign the four feature after carefully compare the on and off resonance spectrum in Fig.5(b). According to the energy diagram of Fe $3d^5$ states[31], the resonantly enhanced feature α is assigned to $e_g \uparrow$ ( $\uparrow$ denotes the majority spin) states of Fe 3d, feature β to the $t_{2g} \uparrow$ states with a weakly hybridized O 2p state (mostly of Fe 3d like character). The energy separation between these two states consistent with the energy separation between the unoccupied $e_g \downarrow$ and $t_{2g} \downarrow$ states as shown in Fig.3(b). The feature γ to the hybridization of Fe 3d-O 2p bonding states since the intensity of this feature increased through Fe 2p-3d threshold and reach a maximum at on resonance region.

The above discussed resonant photoemission investigation results show that the whole ranges of $vb3d^n$ bands have mixed with Fe 3d, O 2p and Bi 6p states. In our previous work about $BiFe_{1-x}Mn_xO_3$, also support there have a strong hybridization between the Fe 3d states with O 2p states in pure BFO and the hybridization decreased with increasing Mn content[34]. The resonate photoemission spectra in Fig.5 strongly support our assumption that whole the marked region (α,β,γ and δ) have a Fe 3d characters through hybridization with other orbitals.

## 4. Conclusion

In conclusion, the electronic structure of multiferroic BFO has been investigated by using the PES and XAS. The measured PES survey spectrum, Fe 2p PES and XAS spectra show that the Fe ions are formally trivalent ($3d^5$: $t_{2g}^3\uparrow e_g^2\uparrow$). The measured valence band RPES and O $K$-edge XAS spectra show that the valence band and conduction band mainly consists of Fe $3d$ and O $2p$ states through hybridization, and $d$-$d$ transition between the valance band and conduction band play a key role in electronic properties of BFO.

## Acknowledgements

This study was financially supported by National Natural Science Foundation of China (Grant No.11164026, 61464010, 11375228, 61366001), University Research Project of Xinjiang Uyghur Autonomous Region (XJEDU2014I002) and Doctoral Startup Foundation of Xinjiang University (BS130110). The authors also thank beamlines 4B9B and 4B7B of BSRF for providing the beam time. R. Wu acknowledge supports under the projects 11204303 by NSFC.

**Figure Caption**

Fig.1. PES survey spectrum of BFO films onto a silicon substrate.

Fig.2. PES survey spectrum of (a)O 1s, (b)Fe *2p* and (c) Bi 4f lines of the BFO films. The molar ratio of Bi and Fe was calculated by normalized peak area of Fe Fe $2p_{3/2}$ and Bi $4f_{7/2}$ (see text for details).

Fig.3 Valence band photoelectron and O *1s* X-ray absorption spectrum of the BFO films. The insets at right panel shows an expended image of the XAS spectrum.

Fig.4 Comparison of the Fe 2p XAS of BFO films to those of FeO[29], $Fe_2O_3$[29] and atomic multiplet calculation.

Fig.5 (a) Valence band spectrum of BFO recorded with photon energy across the Fe 2p-3d resonance. (b) On resonance (hv=710.5eV) and off resonance (hv=702.3eV and 717.5eV) PES spectra of BFO.

Fig.6 Schematic expression of (1) direct photoemission and (2), (3) indirect photoemission processes that make major contributions to the measured valence band DOS distribution at around Fermi level upon excitation of a core level electron in threshold region.

Fig.1

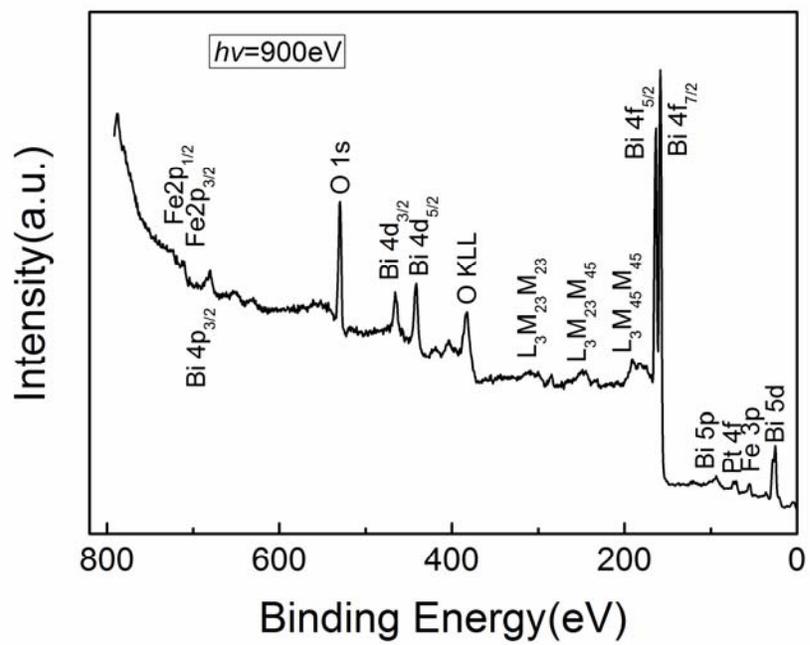

Fig.2

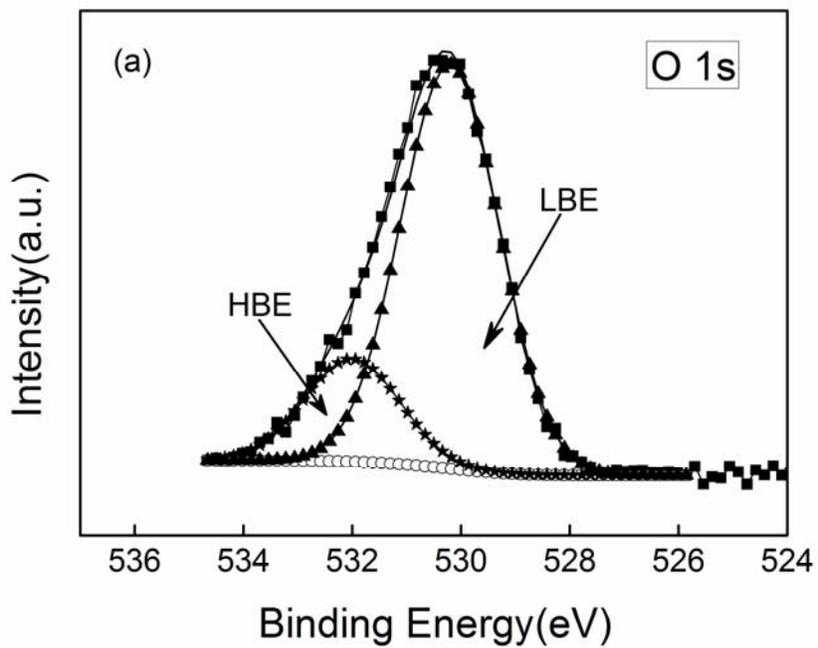

Fig.3

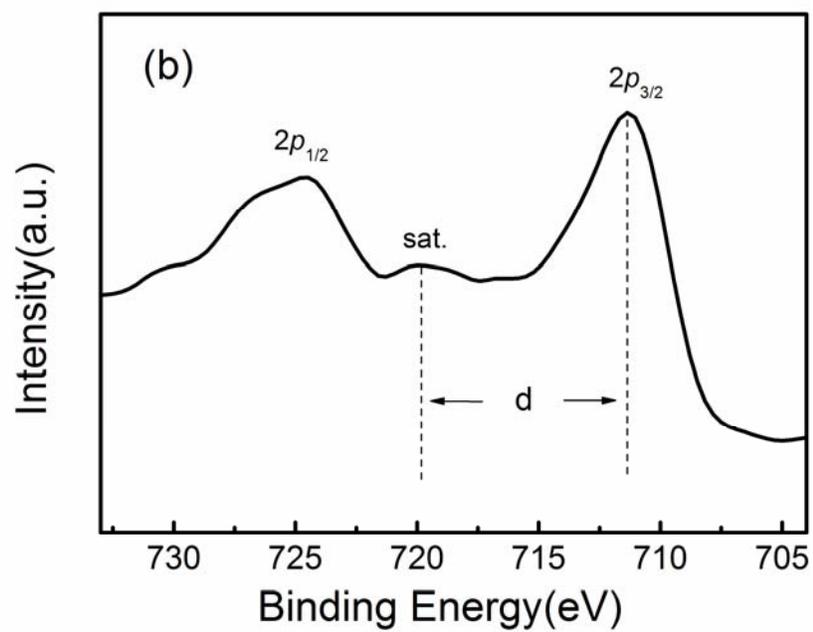

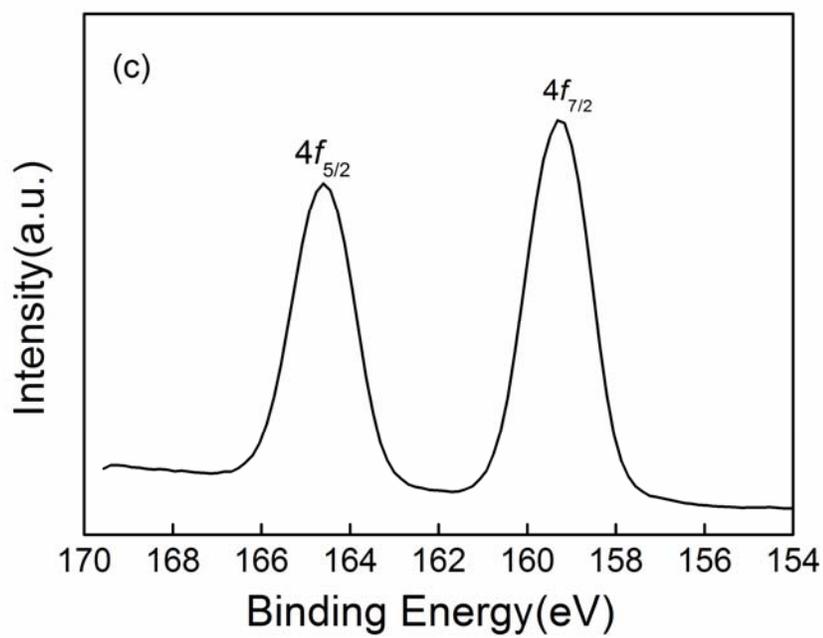

Fig.3

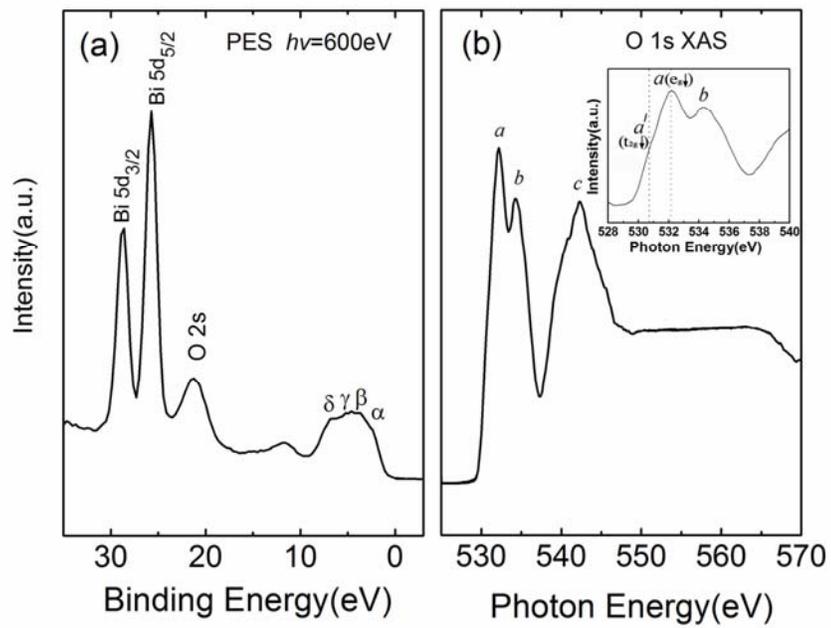

Fig.4

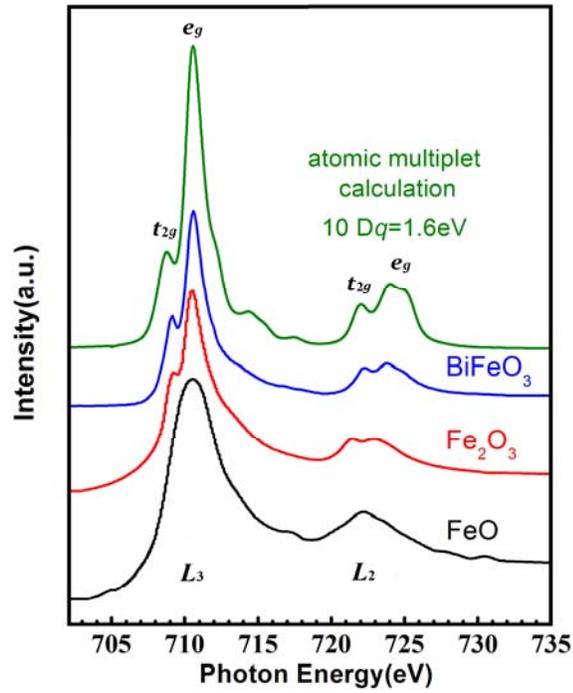

Fig.5 (a) and (b)

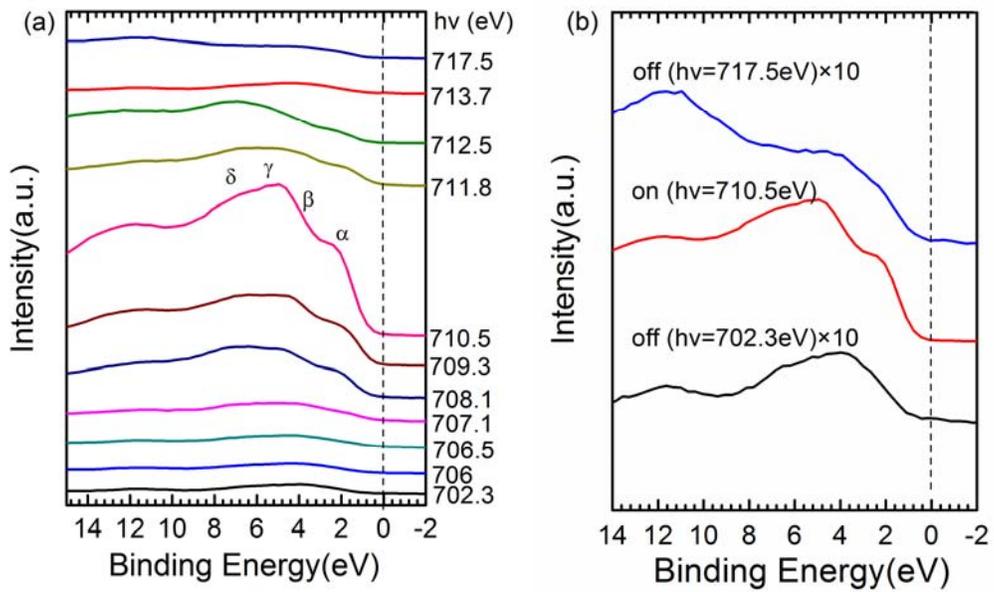

Fig.6

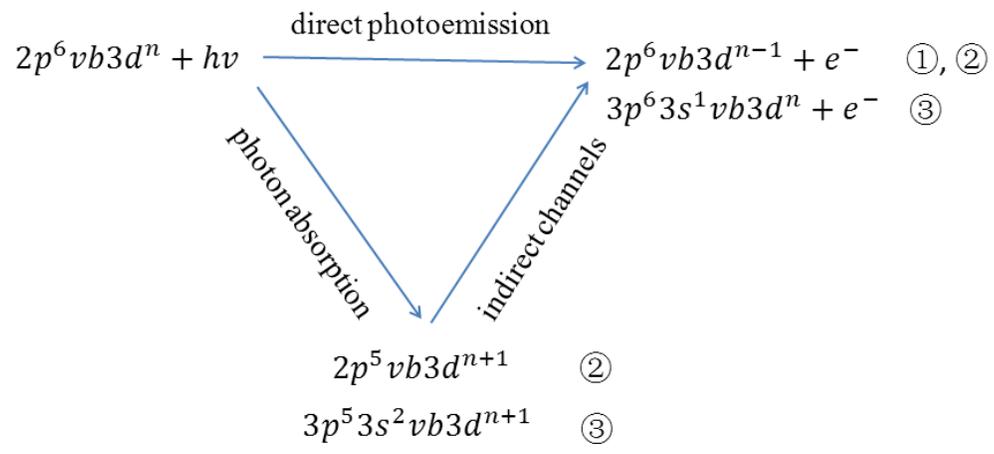